\documentclass{aa}
\usepackage{txfonts}
\usepackage{natbib}
\usepackage{amssymb}
\usepackage{amsmath}
\bibpunct{(}{)}{;}{a}{}{,}
\usepackage{epsf}
\usepackage{graphicx}
\usepackage{epsfig}
\usepackage{graphics}
\usepackage{subfigure}
\usepackage[english]{babel}
\usepackage{rotating}
\usepackage{color}
\usepackage[dvipsnames]{xcolor}

\newcommand{\teff}{$T_{\rm{eff}}$}

\newcommand{\lL}{\ifmmode \log \frac{L}{L_{\sun}} \else $\log \frac{L}{L_{\sun}}$\fi}

\newcommand{\lsun}{\ifmmode L_\sun \else L$_{\sun}$\fi}
\newcommand{\msun}{\ifmmode M_\sun \else M$_{\sun}$\fi}
\newcommand{\mstar}{\ifmmode M_\star \else $M_\star$\fi}

\begin{document}

\title{Impact of a companion and of chromospheric emission on the shape of chromosome maps for globular clusters}
\author{F. Martins\inst{1}
\and J. Morin\inst{1}
\and C.Charbonnel\inst{2,3}
\and C. Lardo\inst{4}
\and W. Chantereau\inst{5}
}
\institute{LUPM, Universit\'e de Montpellier, CNRS, Place Eug\`ene Bataillon, F-34095 Montpellier, France \\
\and
Department of Astronomy, University of Geneva, Chemin des Maillettes 51, 1290, Versoix, Switzerland \\
\and
IRAP, UMR 5277, CNRS and Université de Toulouse, 14, av. E. Belin, 31400, Toulouse, France\\
\and
Laboratoire d'Astrophysique, Ecole Polytechnique Fédérale de Lausanne, Observatoire de Sauverny, 1290 Versoix, Switzerland \\
\and
Astrophysics Research Institute, Liverpool John Moores University, 146 Brownlow Hill, Liverpool L3 5RF, UK
}

\offprints{F. Martins\\ \email{fabrice.martins@umontpellier.fr}}

\date{Received / Accepted }

\abstract
{Globular clusters (GCs) host multiple populations of stars that are well-separated in a photometric diagram - the chromosome map - built from specific Hubble Space Telescope (HST) filters. Stars from different populations feature at various locations on this diagram due to peculiar chemical compositions. Stars of the first population, with field star-like abundances, sometimes show an unexpected extended distribution in the chromosome map.}
{We aim to investigate the role of binaries and chromospheric emission on HST photometry of globular clusters' stars. We quantify their respective effects on the position of stars in the chromosome map, especially among the first population.}
{We computed atmosphere models and synthetic spectra for stars of different chemical compositions, based on isochrones produced by stellar evolution calculations with abundance variations representative of first and second populations in globular clusters. From this we built synthetic chromosome maps for a mixture of stars of different chemical compositions. We subsequently replaced a fraction of stars with binaries, or stars with chromospheric emission, using synthetic spectroscopy. We studied how the position of stars is affected in the chromosome map. }
{Binaries can, in principle, explain the extension of the first population in the chromosome map. However, we find that given the binary fraction reported for globular clusters, the density of stars in the extended part is too small. Another difficulty of the binary explanation is that the shape of the distribution of the first population in the chromosome map is different in clusters with similar binary fractions. Also, the decrease of the binary fraction with radius is not mirrored in the shape of the chromosome map. Additionally, we find that the contribution of chromospheric emission lines to the HST photometry is too small to have an observable impact on the shape of the chromosome map. Continuum chromospheric emission has an effect qualitatively similar to binaries.}
{We conclude that binaries do have an impact on the morphology of the chromosome map of globular clusters, but they are unlikely to explain entirely the shape of the extended distribution of the first population stars. Uncertainties in the properties of continuum chromospheric emission of stars in GCs prevent any quantitative conclusion. Therefore, the origin of the extended first population remains unexplained.}


\authorrunning{Martins et al.}
\titlerunning{Chromosome map in globular clusters}

\maketitle


\section{Introduction}
\label{s_intro}

Globular clusters (GCs) host multiple stellar populations (MSPs) that are identified either photometrically or spectroscopically. In color-magnitude diagrams (CMDs) built with specific filters, they show multiple (almost parallel) sequences from the main sequence to the giant branches \citep[red giant branch -RGB- and asymptotic giant branch -AGB, see e.g., ][]{bedin04,piotto07,soto17}. Spectroscopy indicates that a fraction of stars have chemical compositions similar to field stars, while others show enrichment in nitrogen, sodium, and (sometimes) aluminum together with depletions in carbon, oxygen, and (sometimes) magnesium \citep[e.g.,][]{sneden92,gratton07,Lindetal09,carretta10,marino11,carretta15}. 
Stars with different chemical compositions are found on different sequences of the CMDs \citep[for recent reviews see e.g.,][]{bl18,gratton19}. 

A powerful diagram to separate MSPs through multiband photometry was introduced by \citet{milone15}. Called "chromosome map" thanks to its morphology, it is based on two indices built on a specific combination of Hubble Space Telescope (HST) filters\footnote{\url{http://www.stsci.edu/hst/instrumentation/wfc3}\\ \url{http://www.stsci.edu/hst/instrumentation/acs}}. The first one is the color (m$_{275W}$-m$_{814W}$), where the numbers refer to the HST filters. The second index is the color difference (m$_{275W}$-m$_{336W}$)-(m$_{336W}$-m$_{438W}$), which we refer to as C$_{438W}$ in this paper.
In the (pseudo) CMDs showing, respectively, m$_{814W}$ versus (m$_{275W}$-m$_{814W}$) and m$_{814W}$ versus C$_{438W}$, two lines are defined to bracket the giant branch. The relative position of stars with respect to these two lines is quantified by $\Delta (m_{275W}-m_{814W})$ and $\Delta C_{438W}$ in each CMD \citep[see eqs 1 and 2 of][]{milone17}. The chromosome map shows the latter as a function of the former (see e.g., right panel of Fig.\ \ref{fig_cmd}).  

Several papers \citep{milone17,milone18,milone19,Zennaro19,Saracino19} present a collection of chromosome maps for Galactic and extra-Galactic globular clusters observed by the HST Survey of GCs \citep{piotto15,nardiello18} and follow-up studies. The general shape is a cloud of points stretching from the bottom right to the upper left of the diagram. Stars of the first population (i.e., with field-like surface abundances) lie at the origin - coordinate (0,0) - while stars of the second population - with peculiar abundances - are located on the rest of the observed sequence. In this paper, we refer to these populations and sequences as P1 and P2, respectively, as is commonly done \citep[e.g.,][]{lardo18}. Spectroscopic analysis of stars along the chromosome maps of several GCs confirms that P1 corresponds to stars that have halo-like abundances, while P2 stars align along the well-known C-N and O-Na anticorrelations  \citep{cz19,marino19a}.

In the chromosome map, the Y axis is sensitive to the shape of the spectral energy distribution (SED) in the UV and optical blue. This wavelength region contains molecular lines from CN, CH, NH, SiO, and OH and thus depends on the abundance of carbon, nitrogen, and oxygen \citep{sbordone11,dotter15}. In fact, given the relative changes of these three elements among various subpopulations of GCs, it is the nitrogen content that dominates $\Delta C_{438W}$ \citep[see Fig.~8 of][]{milone18}.
The X axis, $\Delta (m_{275W}-m_{814W})$, probes the SED between the UV and the near-infrared. At first order, it may be an indicator of effective temperature. In that case, its variation can be attributed to a modification of the helium content, as suggested by \citet{milone18} and \citet{lardo18}. Indeed, a higher helium mass fraction implies a lower opacity, which in turn makes a star of a given mass, age, and metallicity hotter \citep[e.g.,][]{wil16}. Metallicity can also affect the SED, with more metal-rich stars being cooler \citep{marino19}.
The dependence of $\Delta m_{275W}-m_{814W}$ and $\Delta C_{438W}$ on light element abundances naturally explains the shape of the chromosome map from the origin through the P2 sequence, since second population stars are expected to be polluted in products of the CNO cycle, meaning they should be N and He rich \citep[e.g.,][and references therein]{PCI17}.

A more surprising feature is the extension of the P1 sequence mainly along the X axis but with a small tilt compared it. This elongation is observed in some GCs \citep[see e.g., Fig.~1 of][]{marino19a}. In view of the dependence on surface abundances shown by \citet{milone18}, these authors, as well as \citet{lardo18}, attributed the existence of an extended P1 to a variation of the helium content among first population stars. However, this variation must not be accompanied by any nitrogen enrichment to prevent a too large increase of $\Delta C_{438W}$. Such a trend - helium enrichment, but with primordial nitrogen - is at odds with the current paradigm according to which hydrogen burning through the CNO cycle (and the NeNa-chain) is responsible for the observed chemical patterns \citep[e.g., ][]{cc16,bl18,gratton19}. Additionally, \citet{tailo19} argue that a different helium content among the first population leads to inconsistent properties of horizontal branch and RR Lyrae stars. 

The shape of the SED, probed by the (m$_{275W}$-m$_{814W}$) color, may change under other effects than variations of effective temperature. In the UV range stars on the RGB and AGB, which are used to build the chromosome map, emit relatively little flux. Any additional source of photons at those wavelengths may thus alter the magnitude in the bluest filters. In the present paper, we investigate how the presence of a companion and chromospheric emission impact the shape of the chromosome map, and especially the extension of the P1 sequence. In Sect.\ \ref{s_clu}, we describe how we build synthetic GCs with multiple populations. We then study the effects of binaries (Sect.\ \ref{s_bin}) and chromospheric emission (Sect.\ \ref{s_chromo}). We summarize our results in Sect.\ \ref{s_conc}. 

\begin{figure}[t]
\centering
\includegraphics[width=0.49\textwidth]{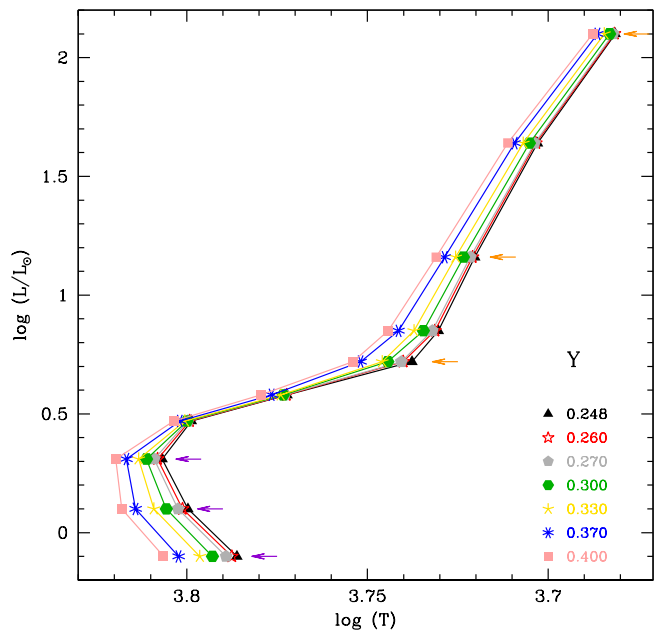}
\caption{Hertzsprung-Russell diagram (HRD) showing isochrones computed by \citet{wil15} for a metallicity of [Fe/H]=-1.53. Different colors correspond to different chemical compositions, labeled by their initial helium mass fraction. Symbols correspond to positions along the isochrones where atmosphere models and synthetic spectra have been calculated. The arrows indicate the models used for binary combinations.}
\label{fig_hrd}
\end{figure}

\section{Synthetic cluster construction}
\label{s_clu}

\begin{figure*}[t]
\centering
\includegraphics[width=0.49\textwidth]{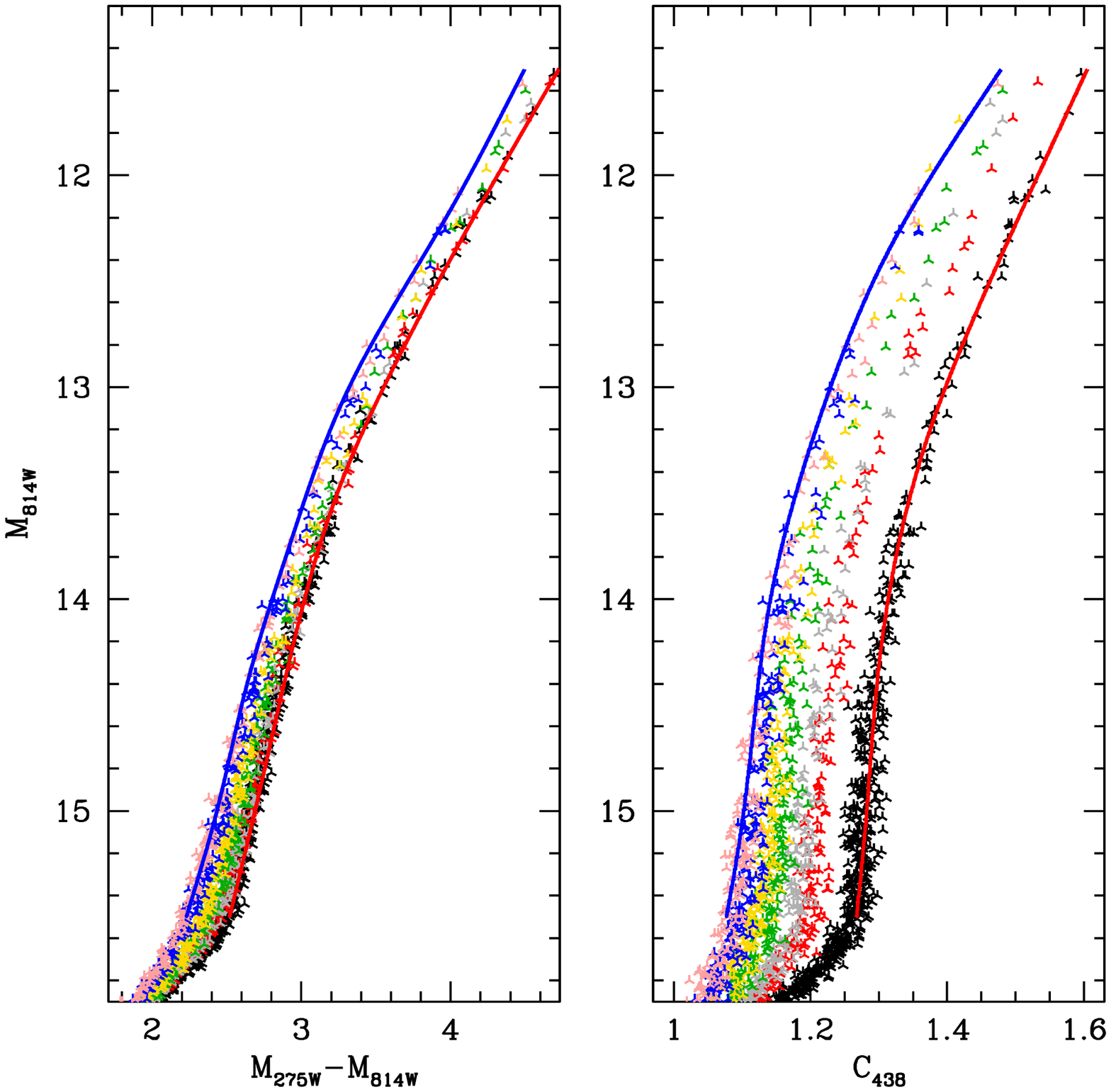}
\includegraphics[width=0.49\textwidth]{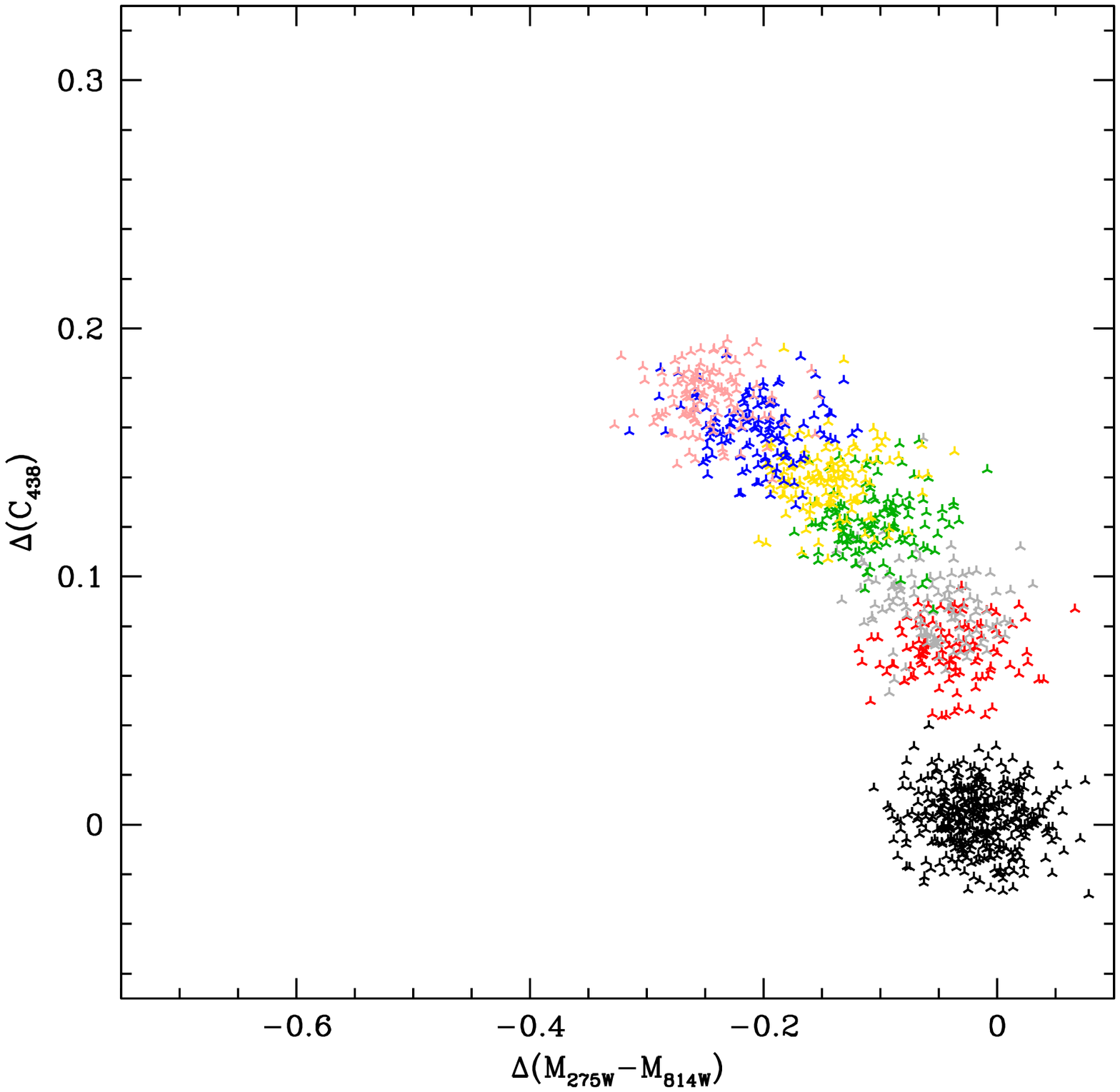}
\caption{Left and middle panels: Magnitude in the F814W filter as a function of the magnitude difference in the F275W and F814W filters (left) and the color index C$_{438W}$ (middle). The symbols denote the stars in the synthetic cluster. The red and blue lines in each panel are the fiducial lines used to calculate color differences in the chromosome map. Right panel: Chromosome map. For clarity, the different populations are color-coded according to their initial chemical composition. Colors are the same as in Fig.\ \ref{fig_hrd}.}
\label{fig_cmd}
\end{figure*}

We built synthetic CMDs from which we have extracted the chromosome map. To do so, we started from the stellar evolution models and isochrones presented by \citet{wil15} for [Fe/H]=-1.53, typical of the GCs NGC~5272, NGC~5986, NGC~6205, NGC~6254, NGC~6584, NGC~6752, and NGC~6981 \citep[see their chromosome maps in Fig.~5 of][]{milone17}. 
The initial compositions assumed for the second population models and isochrones rely on predictions of the fast-rotating massive star scenario \citep[FRMS;][]{Decressin07b,Decressin07a,Krause13}.
In particular, they extend up to very high values of the initial He mass fraction (Y varies from 0.248, the value assumed for P1, up to 0.8, the value corresponding to pollution by FRMS at the end of the main sequence). We note, however, that in these models, the abundance variations of the light elements relevant for our study already occur for very limited He variations. In particular, N increases by 0.5 dex for Y=0.260, and for Y=0.4 the nitrogen enhancement is already of about 1~dex. However, these enrichments (N for a given Y) are not as fast as for other scenarios (e.g., in the case of pollution by supermassive stars' nitrogen enrichment reaches $\sim 1.3$~dex for Y=0.38; \citealt{DenissenkovHartwick14,Gieles18}). In Fig.\ \ref{fig_hrd}, the black isochrone is the nonpolluted (P1) case, while other colored lines correspond to different degrees of chemical pollution. For simplicity, we label them according to their initial helium content, but other elements are also changed: the higher the helium content, the higher (lower) the nitrogen (carbon and oxygen) abundances, among others. We refer to \citet{wil15,wil16} for further details on the evolutionary and isochrone computations. In view of the determinations of the helium mass fraction in GCs, \citet{milone18}, we only considered the isochrones with Y up to 0.4 for this paper.

To move from the HRD to CMDs, we computed atmosphere models and synthetic spectra along the isochrones of Fig.\ \ref{fig_hrd}. We selected points along these isochrones (symbols in Fig.\ \ref{fig_hrd}). At these locations, we adopted the stellar parameters of the evolutionary calculations - \teff, surface gravity, surface abundances - and used them as input for atmosphere models. We used the codes ATLAS \citep{kur14} and SYNTHE \citep{kur05} for atmosphere models and synthetic spectra computations, respectively. The resulting SEDs thus have parameters fully consistent with evolutionary calculations, especially as they have the same abundances.

Once obtained, the SEDs were used to compute synthetic photometry in the HST WFC3 F275W, F336W, F438W, and ACS F814W filters. We could thus build the CMDs m$_{814W}$ versus (m$_{275W}$-m$_{814W}$) and m$_{814W}$ versus C$_{438W}$. In each of these diagrams, we could thus place stars selected by their positions in the HRD - meaning selected by their effective temperature and luminosity - according to their corresponding magnitudes/colors. For each chemical composition, we subsequently interpolated between the points in the two CMDs to build a synthetic isochrone. We used an extinction E(B-V)=0.60 and a distance modulus of 13.15 to set the magnitude scale. These choices are tailored to reproduce the parameters of the cluster NGC~6752 but we stress that the color differences used to build the chromosome map are independent of them (assuming homogeneous extinction among cluster stars).

The next step consisted of the simulation of a synthetic cluster with different populations, such as stars with different chemical compositions. To do so, we first draw points at random along each isochrone, in the two CMDs, assuming the same distribution of m$_{F814W}$ magnitudes as that of the cluster NGC~6752. For each randomly selected m$_{F814W}$, the color (either (m$_{275W}$-m$_{814W}$) or C$_{438W}$) was read from the synthetic isochrone. For a realistic study, we introduced in each color a dispersion drawn randomly from a Gaussian distribution centered on the theoretical color, and with a dispersion equal to 1/3$^{rd}$ of the color dispersion computed by \citet{martins18}. This choice was made to obtain a dispersion in the chromosome qualitatively consistent with observations. The final step consisted of the selection of populations from different chemical compositions. For this, we adopted the following fractions of populations: 34\% with Y=0.248, 11\% with Y=0.260, 11\% with Y=0.270, 11\% with Y=0.300, 11\% with Y=0.330, 11\% with Y=0.370, and 11\% with Y=0.400. As such, the cluster is made of 1/3 of stars from the first population and 2/3 from the second one, a classic ratio among GCs \citep{bl18} and the actual value for NGC~6752 \citep{milone17}. We created a synthetic cluster with a total of 17,000 stars. This number is tailored to have about the same number of stars on the RGB+AGB as in NGC~6752. The final synthetic cluster in the two CMDs is shown in the left and middle panels of Fig.\ \ref{fig_cmd}. The different populations are still visible (especially in the  m$_{814W}$ versus C$_{438W}$ diagram) in spite of the dispersion introduced along each isochrone.

The chromosome map was subsequently built from the two CMDs following the method described by \citet{milone17}. The only difference is that we selected the so-called fiducial lines visually rather than using number counts in different magnitude bins. The fiducial lines are shown by the red and blue solid lines in Fig.\ \ref{fig_cmd}. 
The resulting chromosome map clearly shows the groups of stars with different chemical compositions. The red and black populations, which correspond to the least and nonchemically processed ones, respectively, are separated by a region almost devoid of stars, and are located almost on top of each other. This is explained by the rapid increase in nitrogen (which mainly dominates the C$_{438W}$ index) and the slowest helium enrichment in the early stages of the CNO cycle. The material that polluted the red population is made of such nitrogen-rich and helium (quasi) normal material.

\begin{figure}[t]
\centering
\includegraphics[width=0.49\textwidth]{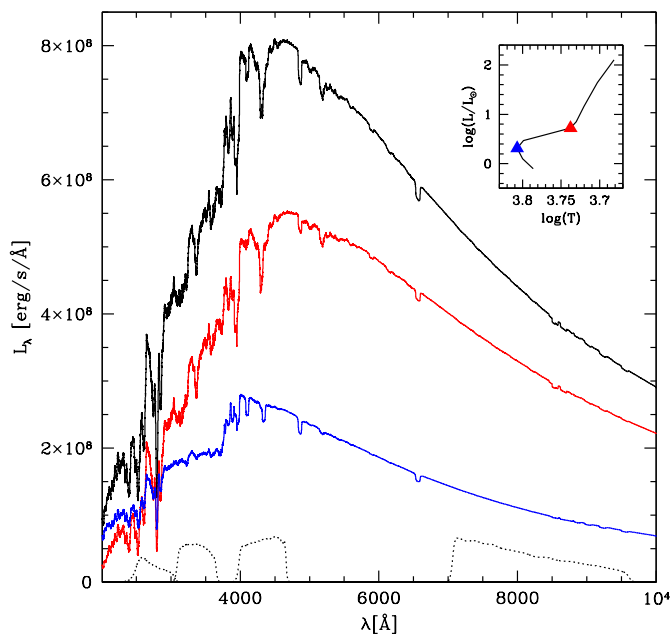}
\caption{Composite spectrum (black line) made of one low-luminosity RGB (red line) and one turn-off (blue) star of the Y=0.248 isochrone. The HRD in the insert shows the position of the two stars (filled triangles) along the Y=0.248 isochrone. The dotted lines show the HST filters throughputs (F275W, F336W, F438W, and F814W).}
\label{fig_specbin}
\end{figure}

\section{Inclusion of binaries\\ }
\label{s_bin}

We now proceed with the estimate of the impact of binaries on the distribution of stars in the chromosome map. We first describe our method and subsequently discuss our results.

\subsection{Method}
\label{s_bin_meth}

To investigate the role of binaries on the chromosome map, we first studied the impact of a companion on the magnitude m$_{814W}$ and the colors (m$_{275W}$-m$_{814W}$) and C$_{438W}$. For that, we proceeded as follows. Firstly, we selected three representative spectra of models along the Y=0.248 isochrone: the most luminous one, the middle one, and the one at the bottom of the RGB (see orange arrows in Fig.\ \ref{fig_hrd}). For each of these models, we added the spectra of stars on the main sequence, from the turn-off to the least luminous one in Fig.\ \ref{fig_hrd} (purple arrows). We thus assumed that each system is made of two stars of the same population. Fig.\ \ref{fig_specbin} illustrates the process: the red line is the RGB spectrum, to which we added the main sequence blue spectrum to finally obtain the total black spectrum. The latter spectrum is used to compute synthetic photometry, from which we calculated the magnitude and color differences compared to the initial RGB model. For each RGB model, we repeated the process with the three companion's main-sequence stars, ending up with nine combinations of spectra and the associated color differences. 
We performed the same exercise with the RGB and main-sequence stars on the Y=0.300 isochrone to check the effects of chemical composition on binary colors but found little differences compared to the Y=0.248 case. 

\begin{figure}[t]
\centering
\includegraphics[width=0.49\textwidth]{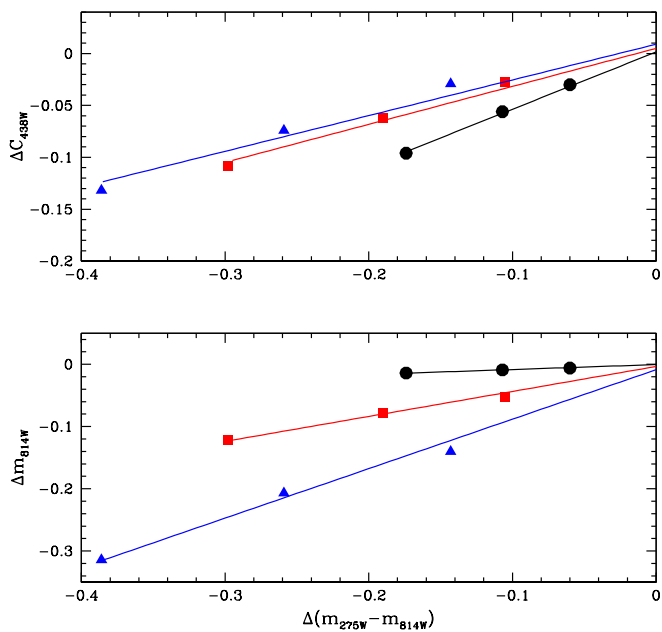}
\caption{Relation between color correction $\Delta C_{438W}$ and $\Delta (m_{275W}-m_{814W})$ (top panel) and $\Delta m_{814W}$ and $\Delta (m_{275W}-m_{814W})$ (bottom panel) resulting from binarity for Y=0.248 stars. Each symbol corresponds to a combination of spectra as described in Sect.\ \ref{s_bin_meth} and shown in Fig.~\ref{fig_specbin}. Each color refers to a given giant star (black: the most luminous giant; red: the intermediate luminosity giant; blue: the giant at the bottom of the RGB), and thus to a given magnitude bin defined in Sect.\ \ref{s_bin_meth}. Linear regressions to each set of colored points are shown. For the linear regressions, we also added the point at (0,0) to ensure that all color corrections are negligible in case of low-mass companions. See also, Sect.\ \ref{s_massrat}.}
\label{fig_fitdelta}
\end{figure}

We then started from the synthetic cluster built in Sect.\ \ref{s_clu}. We divided the RGB into three magnitude bins: $m_{814W}>14.5$, $13.0<m_{814W}<14.5,$ and $m_{814W}<13.0$. 
We randomly selected points from the synthetic cluster. For each selected star, its $m_{814W}$ magnitude fell in one of the three bins defined above.
We estimated a color correction $\Delta (m_{275W}-m_{814W})$ by drawing a value from a Gaussian distribution with a dispersion equal to 0.42 (see Sect.\ \ref{s_massrat} for a discussion of that choice) and multiplying it by $\Delta (m_{275W}-m_{814W})_{max}$. The latter is the maximum correction possible in each of the three m$_{814W}$ bins. It corresponds to the abscissa of the point to the left of each line in Fig.\ \ref{fig_fitdelta}. We thus obtained a color correction $\Delta (m_{275W}-m_{814W})$ for the selected star. We subsequently read the corresponding correction on $\Delta C_{438W}$ and $\Delta m_{814W}$ from Fig.\ \ref{fig_fitdelta}. For the bin with the brightest stars ($m_{814W} < 13.0$), we used the relations shown in black. For the $13.0 < m_{814W} < 14.5$ bin (respectively $m_{814W} > 14.5$ bin), we used the relations shown in red (blue). Once obtained, the color corrections were added to the photometry of the initially single giant star. We repeated the process until 10\% of the stars were replaced by binaries.

\begin{figure*}[ht]
\centering
\includegraphics[width=0.49\textwidth]{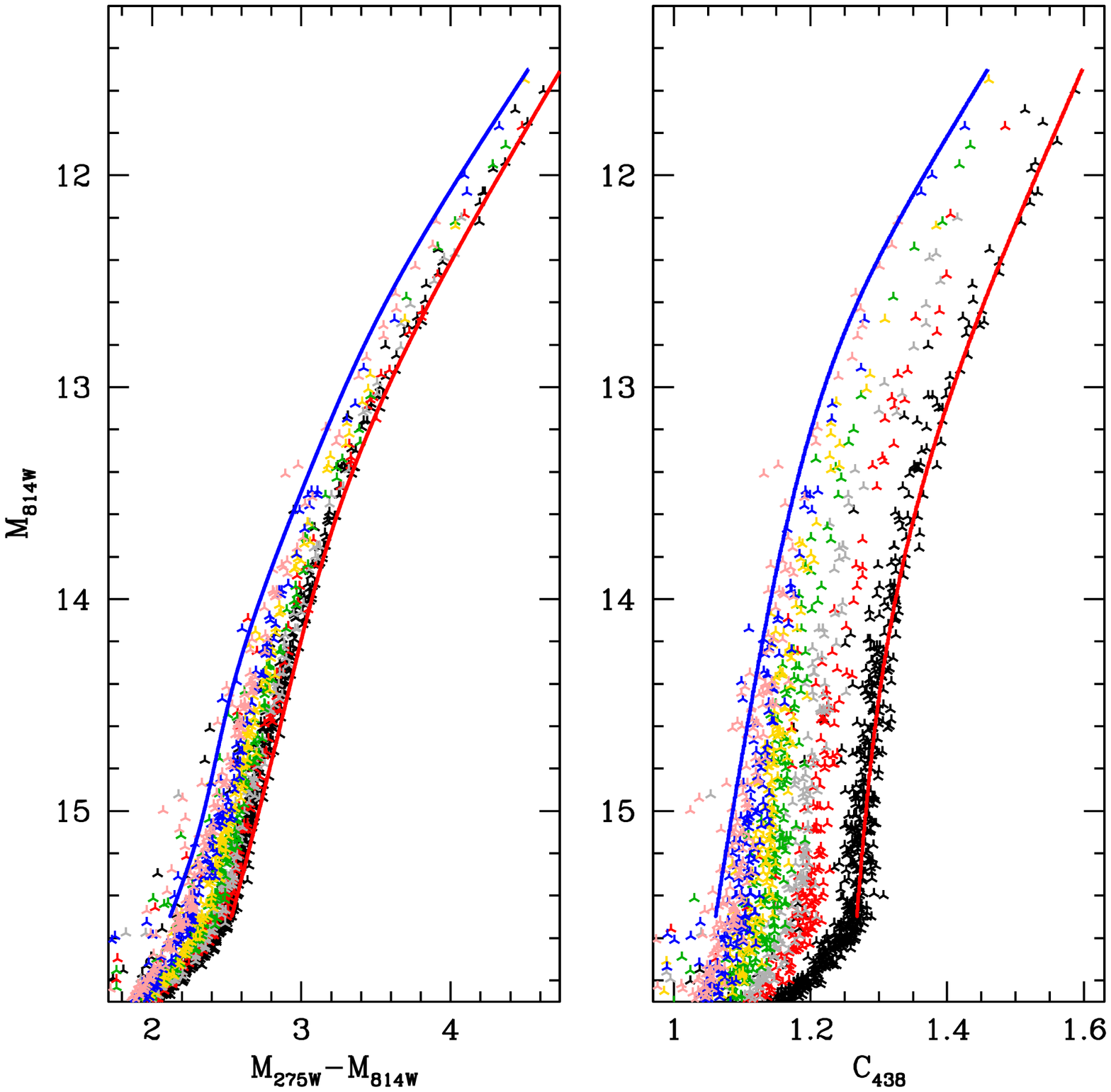}
\includegraphics[width=0.49\textwidth]{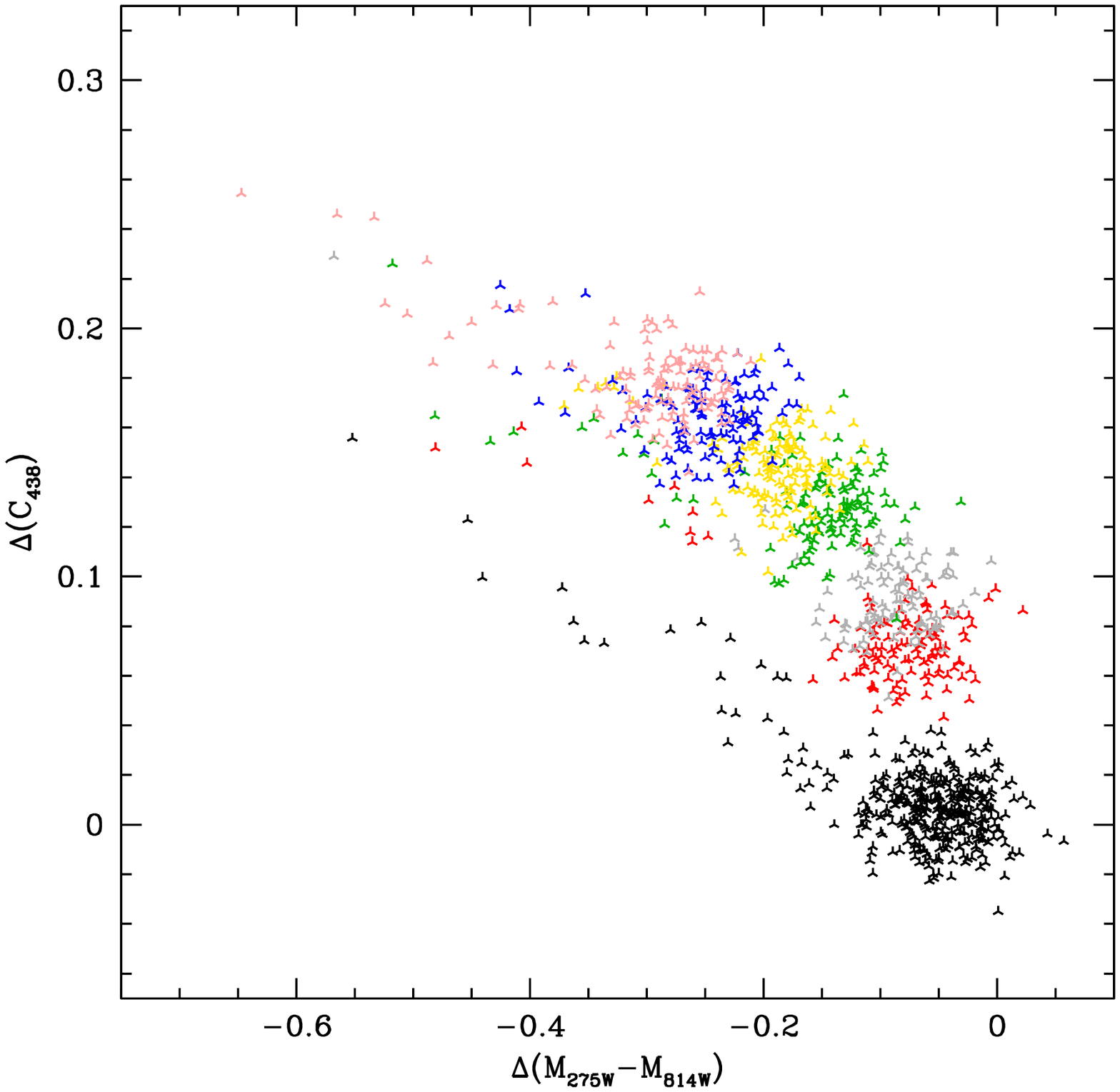}
\caption{Same as Fig.\ \ref{fig_cmd} now including 10\% of binaries.}
\label{fig_cmdbin}
\end{figure*}

\begin{figure*}[ht]
\centering
\includegraphics[width=0.49\textwidth]{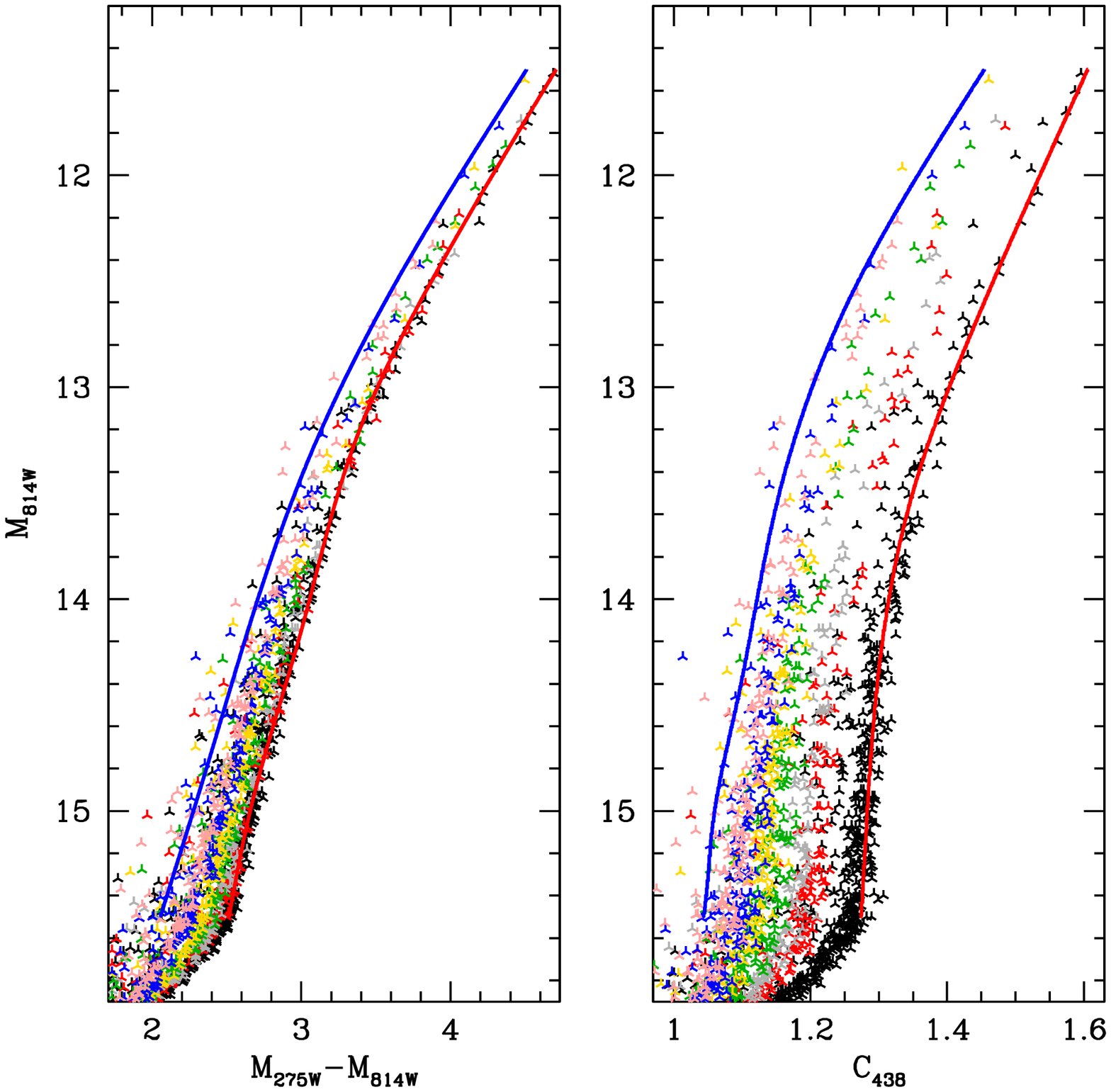}
\includegraphics[width=0.49\textwidth]{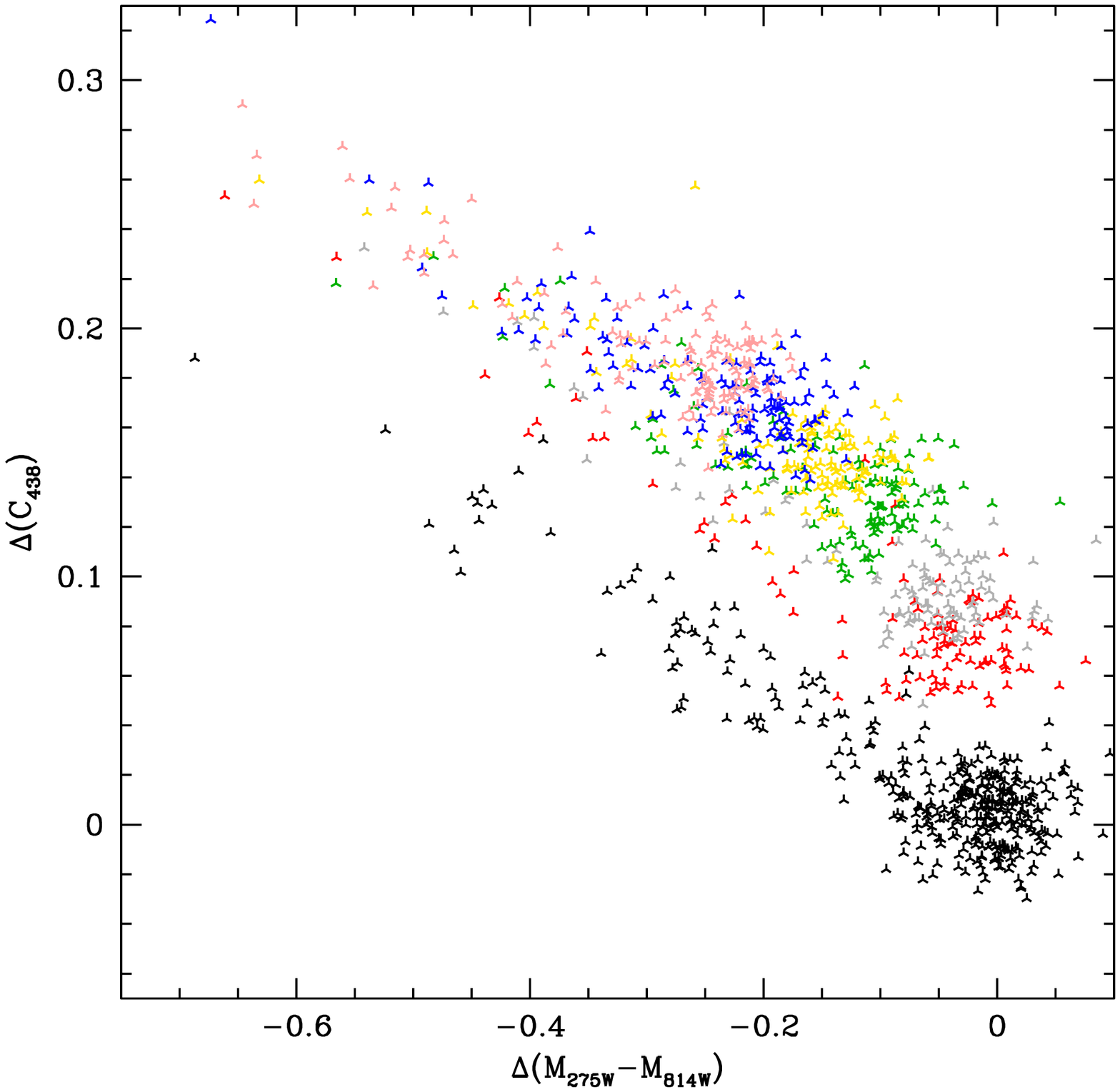}
\caption{Same as Fig.\ \ref{fig_cmdbin} with 30\% of binaries.}
\label{fig_cmdbin0p3}
\end{figure*}

The final step consisted of building the CMDs and chromosome map from the resulting cluster that now contains 10\% of stars in binary systems. The results are shown in Fig.\ \ref{fig_cmdbin}. The color corrections of the Y=0.248, 0.260 and 0.270 populations are assumed to be those of the Y=0.248 binary combinations. For the other populations (Y$\geq$0.300), we used the corrections obtained for the binaries of the Y=0.300 population. In practice, the corrections resulting from binarity depend little on the chemical composition. We also considered a binary fraction of 30\%, which corresponds to the highest values reported for GCs \citep[e.g.,][]{milone16}. The results are shown in Fig.\ \ref{fig_cmdbin0p3}.

\subsection{Do binaries explain the P1 extension?}
\label{s_bin_disc}

Inspection of the synthetic chromosome maps in Figs.\ \ref{fig_cmd},  \ref{fig_cmdbin}, and \ref{fig_cmdbin0p3} reveals that binaries impact the shape of P1. Initially centered around the (0,0) coordinates point, this population is elongated towards the left and upper part of the diagram because of binaries. The direction of the main axis is tilted by $\sim$14$^o$ compared to the X axis. This is qualitatively in agreement with the 18$^o$ measured by \citet{milone17} for the cluster NGC~6723. Binaries are therefore a possible explanation for the extension of the P1 sequence. This was also recently noted by \citet{marino19}. 
However, our simulations indicate that the density in the elongated part of P1 depends on the binary fraction. What are the observational values?

\citet{sollima07} determined the minimum binary fraction in 13 GCs from photometry on the main sequence. They obtained values between 6 and 10\%, except for three clusters where the numbers can reach 10 to 20\%. Using various assumptions on the mass ratio distribution, they estimated a total binary fraction in the range 10-20\% (40-65\% for the extreme cases). They also showed that the binary fraction is larger in the core than in the outer parts of the clusters. This result was confirmed by \citet{dal11} for NGC~6254: the binary fraction decreases from 14\% in the cluster's core to 1.5\% in a region located between one and two times the half-light radius. According to \citet{sollima07}, the clusters with the largest binary fractions are the youngest, suggesting an evolution with time \citep[but see][]{milone16}. Using the same method, \citet{milone12} extended this study to 59 GCs and found results consistent with those of \citet{sollima07}. In addition, they showed that the binary fraction in GCs was, on average, lower than in the field, and that the binary fraction was anticorrelated with the cluster's mass. \citet{jb15} determined binary fractions mostly in the range of 3-10\% for the 35 GCs they analyzed. This range is similar to those of \citet{sollima07} and \citet{milone12}, although for a given cluster the binary fraction estimated by the three groups may differ significantly. \citet{jb15} confirmed the radial variation of the binary fraction. Using a different method based on the identification of radial velocity variations, \citet{luc15} reported an average binary fraction of $\sim$2\% with a difference between the first and second populations: 4.9\% in the former, 1.2\% in the latter\footnote{The lower binary fraction compared to photometric studies is explained by the more demanding nature of observations for obtaining spectroscopic data: longer exposure times and need for multiepoch observations.}. This trend - a higher binary fraction among the first population stars compared to the second population stars - was also reported by \citet{dorazi10} and \citet{dal18} with even larger differences (nearly an order of magnitude more binaries in the first population). Finally, using ESO/VLT/MUSE spectroscopy, \citet{giesers19} report a binary fraction of 6.75$\pm$0.72\% in NGC~3201. In a follow-up study of this cluster, \citet{kamann20} find that P1 hosts more binaries than P2 (23.1$\pm$6.2\% versus 8.2$\pm$3.5\%).

\begin{figure*}[t]
\centering
\includegraphics[width=0.49\textwidth]{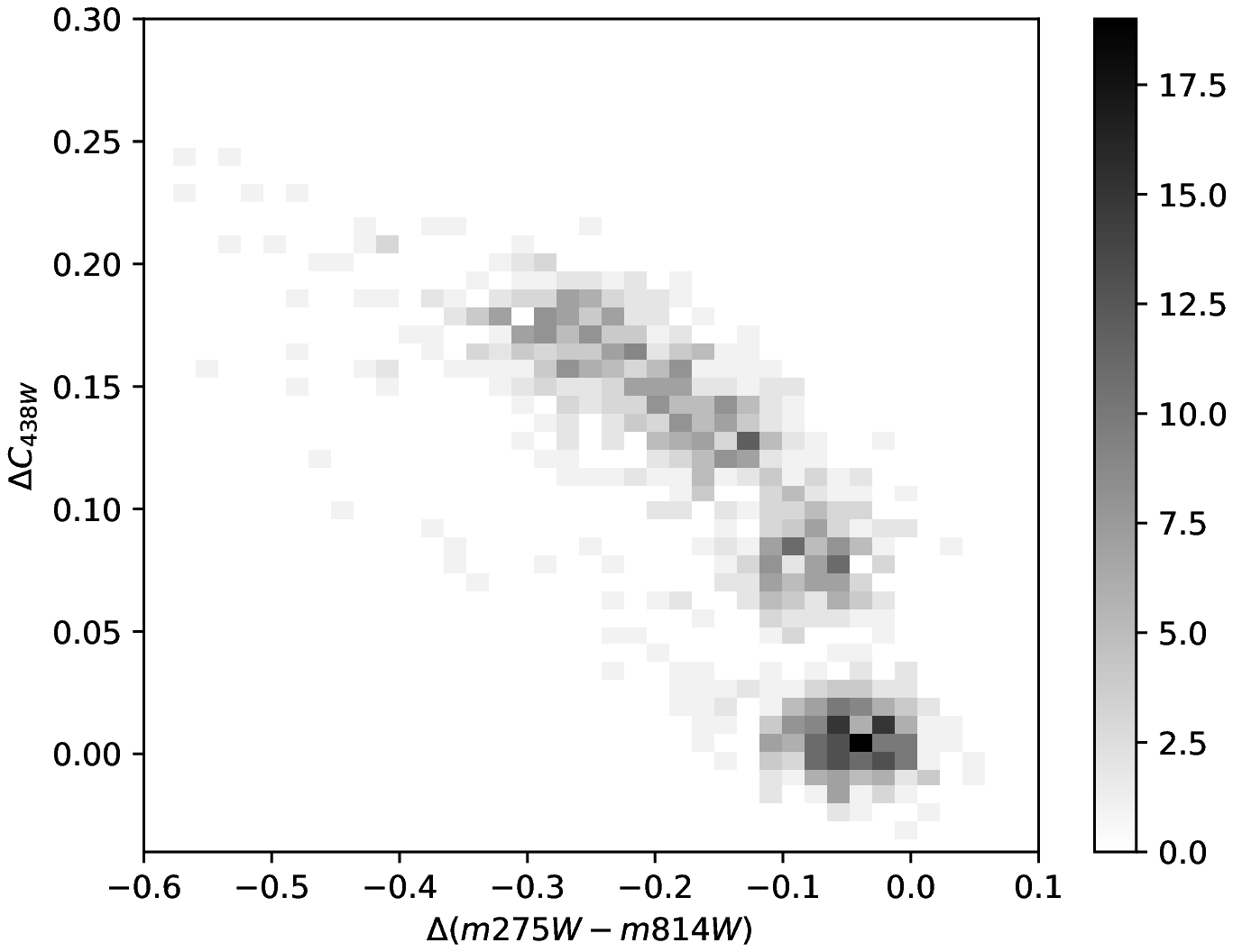}
\includegraphics[width=0.49\textwidth]{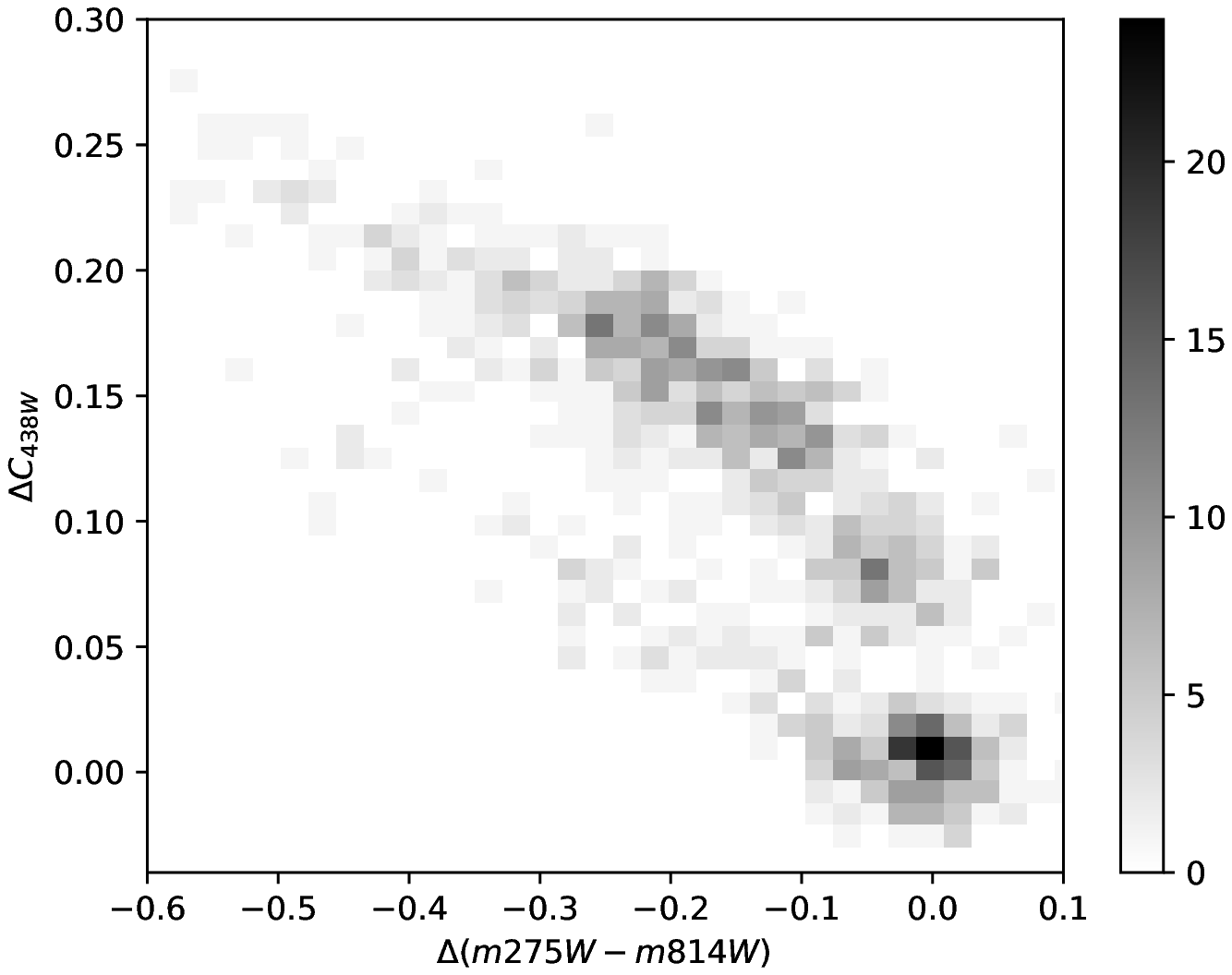}\\
\includegraphics[width=0.49\textwidth]{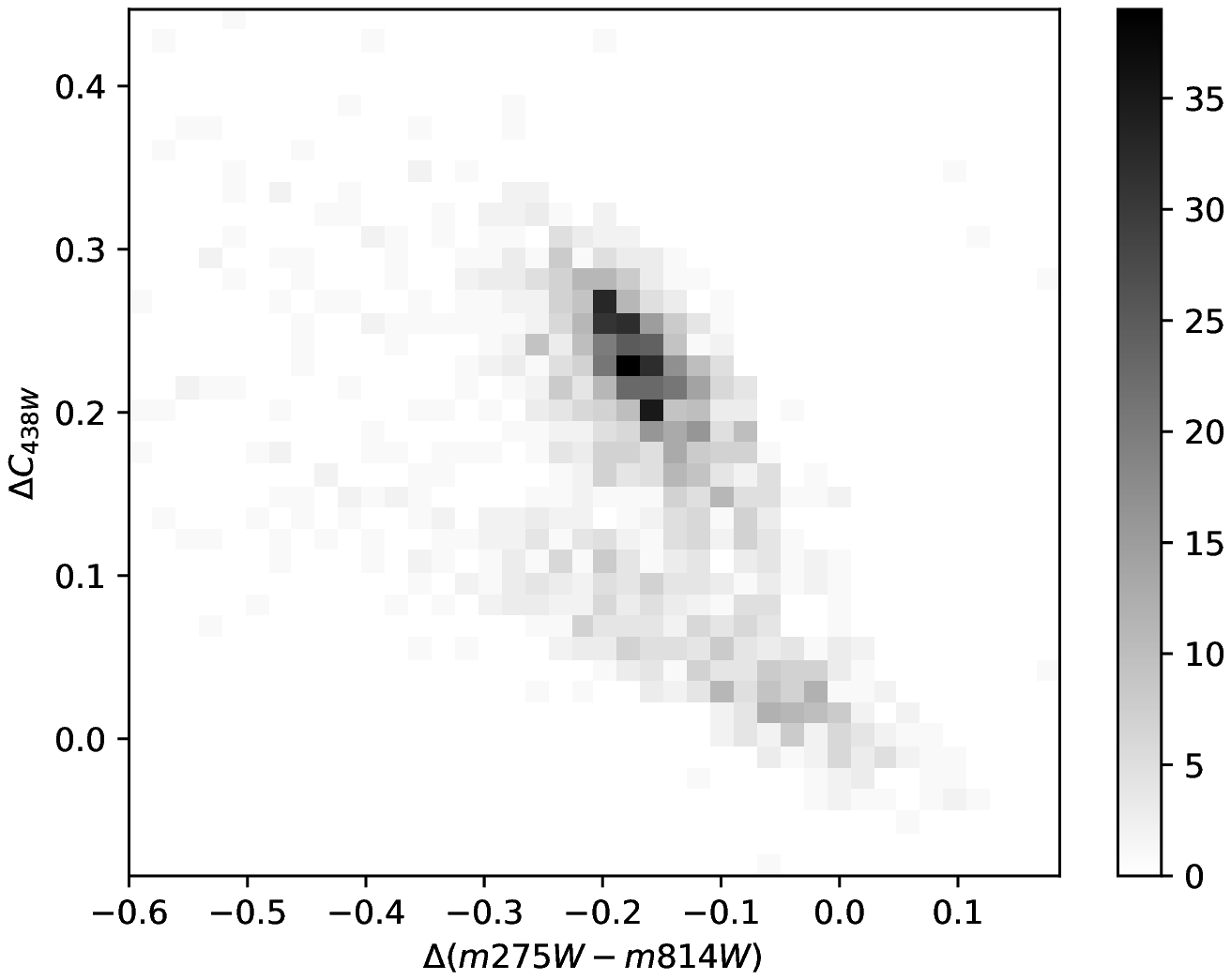}
\caption{Hess diagrams based on chromosome maps of the synthetic clusters with 10\% (top left panel) and 30\% (top right panel) of binaries, together with that of of NGC~5272 (bottom panel).}
\label{fig_hess}
\end{figure*}

In view of the binary fractions listed above, a typical chromosome map including binaries should be relatively close to that of Fig.\ \ref{fig_cmdbin} (i.e., 10\% binaries) - at least regarding the first population. The extent of P1 should be present, but relatively unpopulated. NGC~5272 (M3) is a cluster with a metallicity and age close to those of our synthetic clusters. \citet{sollima07} determined a binary fraction within NGC~5272 between 5 and 9\%, while \citet{milone12} reported 3-5\% of binaries. These values are relatively close to the 10\% we adopted in Fig.\ \ref{fig_cmdbin}. \citet{milone17} determined a fraction of stars in the first population of 30.5\% for NGC~5272, close to the value used to build the synthetic chromosome map (see Sect.\ \ref{s_clu}).
Figure\ \ref{fig_hess} shows the density of stars across the chromosome map in NGC~5272 and the synthetic clusters with 10\% and 30\% of binaries. In the latter ones, the peak density is around the (0,0) point. The binaries contribute to producing an elongated structure with a lower density (which increases with the binary fraction). In NGC~5272\footnote{The data are from the HST UV Globular Cluster Survey and are publicly available at \url{https://archive.stsci.edu/prepds/hugs/}. They are from \citet{piotto15} and \citet{nardiello18}.}, the P1 sequence is relatively homogeneously populated with barely a very small overdensity near the origin. The same trend is observed for NGC~6254 (not shown), from which we conclude that either the binary fractions in GCs are underestimated, or binaries contribute only part of the extension of P1, and another process is at play.

\smallskip

\begin{figure*}[t]
\centering
\includegraphics[width=0.49\textwidth]{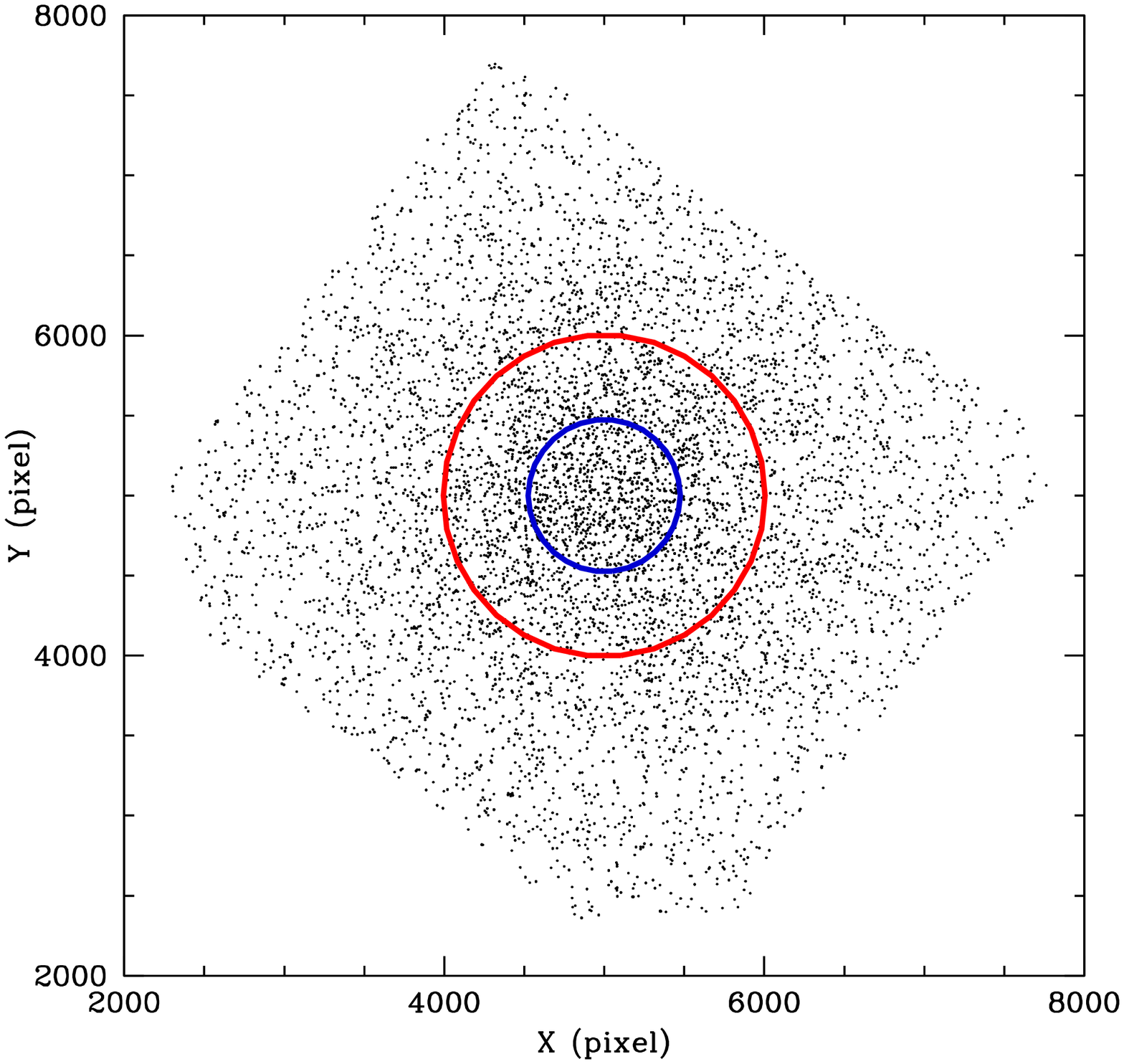}
\includegraphics[width=0.49\textwidth]{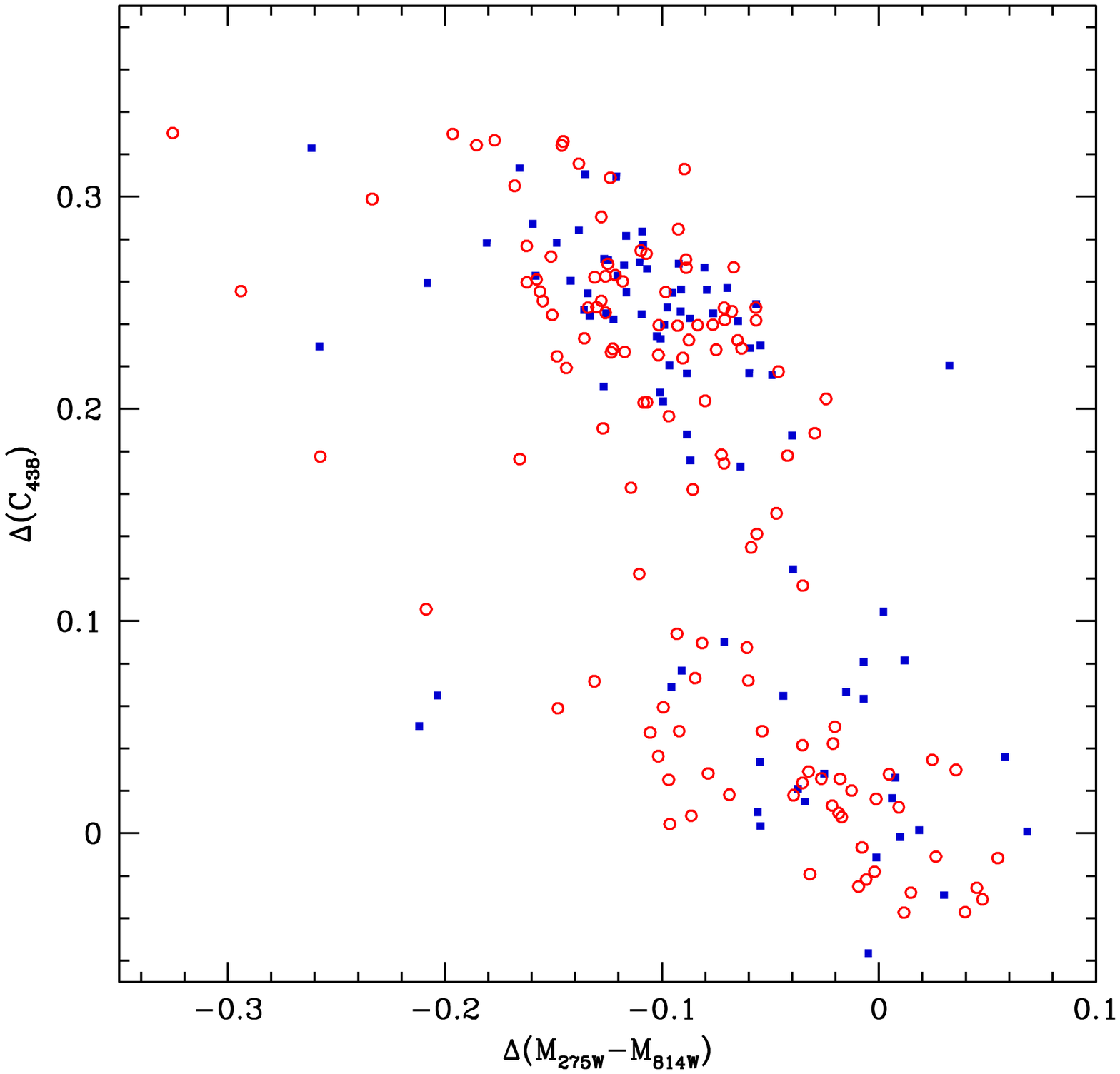}
\caption{Left panel: Position of stars brighter than m$_{814W}$=18.5 in the cluster NGC~6254. The red and blue circles indicate the two regions used to produce the chromosome maps of the right panel. Right panel: Chromosome map of the stars inside the blue circle in the left panel (blue squares) and of the stars in between the red and blue circles (open red circles). Only stars brighter than m$_{814W}$=16.5 (i.e., above the bottom of the RGB) are considered in the chromosome maps. }
\label{fig_6254}
\end{figure*}

A further test of the effects of binaries on the chromosome map is shown in Fig.\ \ref{fig_6254}. We selected the cluster NGC~6254 since; 1) it shows an extended P1 \citep{milone17}, 2) its metallicity corresponds to that of our models, and 3) the binary fraction decreases from the core to the outer parts of the cluster \citep{dal11,milone12}. To investigate whether or not this variation of the binary fraction is reflected in the shape of P1, we selected two regions of the cluster: the core and an annulus around it, as displayed in the left panel of Fig.\ \ref{fig_6254}. We subsequently selected the stars brighter than the bottom of the RGB in each region and constructed their chromosome map. The results, shown in the right panel of Fig.\ \ref{fig_6254}, do not show any significant variation of the extent of the chromosome map. This result does not depend on the choice of the core and annulus regions. If binaries governed the elongation of P1, one would expect a smaller extension in the outer regions of the cluster, which is not observed.

If binarity were the main reason for the extension of P1, one would also find a correlation between binary fraction and P1 extension. 
A simple comparison of two GCs with similar properties, NGC~5272 (M3) and NGC~6205 (M13), shows some qualitative trends. Both clusters are of about the same age, metallicity and mass \citep{har96,mf09}, but M3 shows an extended P1, while M13 does not - see Fig.~5 of \citet{milone17}. According to \citet{milone12}, the total binary fraction in M3 (M13) is between 3 and 6\% (2 and 10\%). Thus, M3 does not have a higher binary fraction that would explain its more extended P1.

\subsection{Effect of the binary mass ratio distribution}
\label{s_massrat}

In Sect.\ \ref{s_bin_meth}, we used a Gaussian distribution with dispersion equal to 0.42 to select color corrections due to binaries. This choice is motivated by the shape of the distribution of the mass of companions to stars in the solar neighborhood according to \citet{dm91}. This distribution is best reproduced by a Gaussian function with the same dispersion. In our combinations of spectra made to estimate the effects of a companion on the photometry of giant stars (Sect.\ \ref{s_bin_meth}), we used the models identified by the arrows in Fig.\ \ref{fig_hrd}. The various combinations produce systems with mass ratios between 0.97 and 0.89 since we only include main-sequence stars relatively close to the turn-off. The color corrections caused by a low-mass component are negligible since; 1) the companion is much fainter, and 2) it is cool. Consequently, the spectrum of the companion barely affects the SED of the giant star. The color corrections resulting from companions with masses lower than 0.89 $\times$ the mass of the giant star are thus extrapolated from the combinations of spectra with the most massive main-sequence stars. They are assumed to follow the linear fits shown in Fig.\ \ref{fig_fitdelta}. When selecting color corrections with a Gaussian distribution, we thus assume that such corrections follow the same distribution as the mass ratio distribution in binaries.

Different assumptions can be made. For instance, \citet{marino19} studied the chemical composition of P1 stars in NGC~3201. They found that two stars located at the extreme left of the sequence showed radial velocity variations, and thus concluded that they were binaries. This prompted them to study the effect of binarity on the shape of the chromosome map in a way similar to that presented here. They concluded that binaries with mass ratios larger than 0.8 could explain the shape of P1. 
They also suggested that binaries may be present mainly in the first population of stars since no extension of P2 is observed. We qualitatively reach the same conclusion regarding P1: binaries may explain part of its elongated shape. However, we stress that given the current knowledge of the binary fraction in GCs, binaries are probably not numerous enough to sufficiently populate the extended P1 sequence. 

\smallskip

\begin{figure}[ht]
\centering
\includegraphics[width=0.49\textwidth]{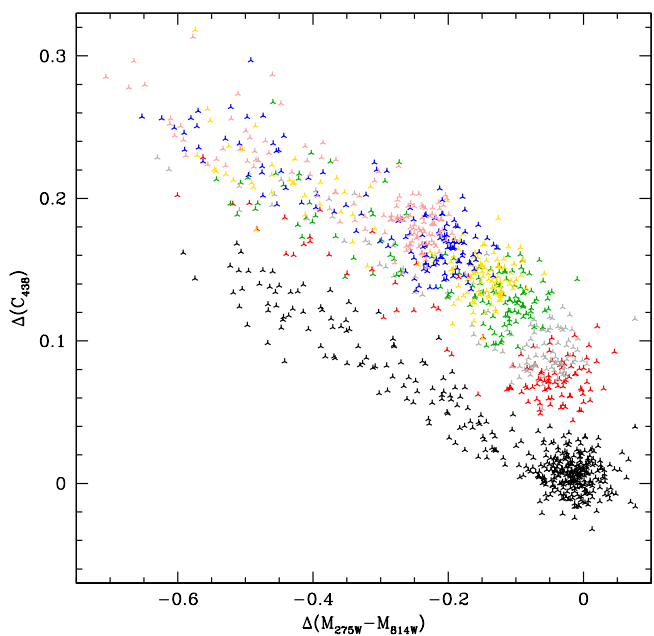}
\caption{Same as chromosome map in Fig.\ \ref{fig_cmdbin0p3}, but assuming a flat distribution of mass ratios among binaries.}
\label{chmapflat}
\end{figure}

\citet{marino19} used a flat distribution of mass ratios among binaries, while we rely on the results of \citet{dm91} for the solar neighborhood, meaning a distribution that favors low-mass companions. To test the effect of this assumption on the shape of the chromosome map, in Fig.\ \ref{chmapflat} we present the same cluster simulation as in Fig.\ \ref{fig_cmdbin0p3} (i.e., 30\% of binaries), but using a flat distribution of the color corrections, and thus implicitly of the mass ratios of binaries. The effect is a more uniformly populated extension of P1, together with a larger number of stars with extreme $\Delta (m_{275W}-m_{814W})$. The explanation is as follows. With a flat-mass ratio distribution, the probability of having a companion of almost equal mass is higher. This means that main-sequence companions close to the turn-off are more likely. These companions imply the largest changes in photometry, because they are brighter (hence affect $m_{814W}$) and hotter (hence affect $(m_{275W}-m_{814W})$) than lower mass main-sequence stars. Consequently, they lead to the largest changes in the chromosome map. The distribution of mass ratios among binaries made of solar-type stars in GCs remains widely unknown. \citet{giesers19} provide the first empirical determination of that distribution in NGC~3201. Its shape is qualitatively consistent with that of \citet{dm91}. In the solar neighborhood, building on the study of \citet{dm91}, \citet{halb03} reported a distribution with a broad peak for mass ratios between 0.2 and 0.7, and a second peak for equal-mass binaries in short-period systems (and no second peak for long-period binaries). \citet{rag10} argued for an almost flat distribution, and confirmed the trend that equal-mass binaries are more frequent among short-period systems. 

Recent surveys \citep{badenes18,moe19} report an increase of the overall binary fraction of solar-type stars when metallicity decreases. The binary fraction at [Fe/H]=-1.0 is $40\pm6\%$ and reaches $53\pm12\%$ at [Fe/H]=-3.0 according to \citet{moe19}. This metallicity range corresponds to that of GCs, for which the estimated binary fraction is much lower \citep{milone12,jb15}. Dynamical effects in the dense environments of GCs most likely affect the properties of the binaries they host \citep{heggie75}. It is thus not unlikely that the mass ratio distribution is also affected, and thus different from field stars.

Clearly, the shape of the mass ratio distribution among GC binaries remains a degree of freedom in the construction of synthetic GCs with binaries. However, whatever the choice, the conclusions remain the same: the extended part of P1 remains less populated than the initial sequence at (0,0) in the chromosome map.

\vspace{0.5cm}

In view of the arguments we presented and discussed above, we conclude that binaries may contribute to the extent of P1, but they are likely not the main driver.

\section{Chromospheric emission}
\label{s_chromo}

In this section we describe the impact of chromospheric emission on the SED of stars, and thus on the shape of the chromosome map. 

\subsection{Line emission}

The atmosphere of solar-type and giant stars is the locus of complex phenomena resulting from the magnetic field, chromosphere, coronae, and winds. This results in stellar activity, the level of which depends on the type of stars. Among the observed phenomena, emission in specific lines is caused by the presence of a chromosphere \citep{schri95}. The magnesium and calcium HK lines in the UV, at 2976-2803~\AA\ and 3934-3968~\AA,\ respectively, are notably concerned. The calcium doublet is a classical indicator of stellar activity \citep{linsky79,noyes84,bal95,wright04,marsden14}.
The \ion{Mg}{ii} lines are located close to the center of the F275W filter, while the \ion{Ca}{ii} lines are on the blue side of the F438W filter. These lines may thus affect photometry in both filters, and consequently the position of stars in CMDs and the chromosome map.

\begin{figure*}[ht]
\centering
\includegraphics[width=0.49\textwidth]{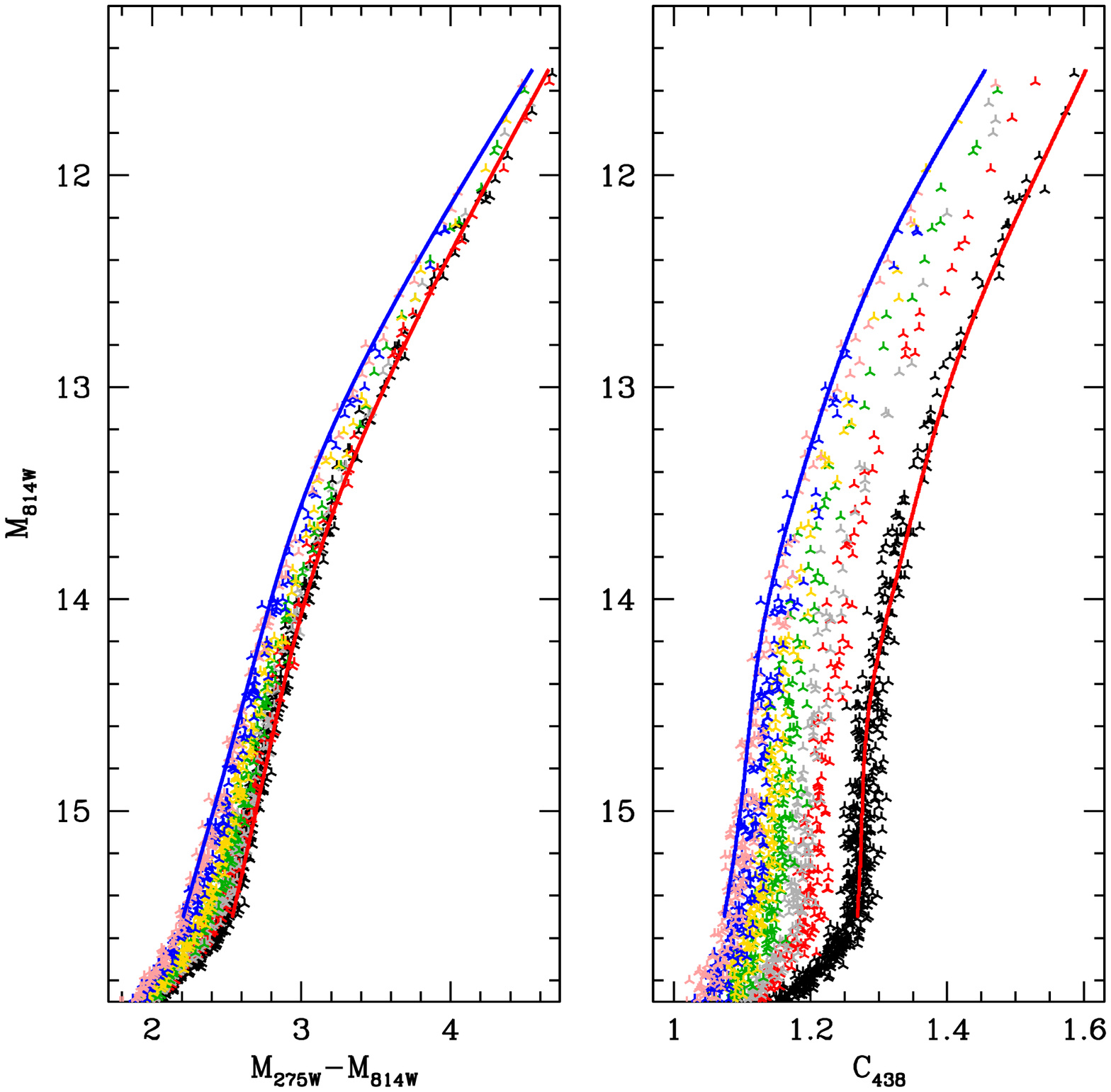}
\includegraphics[width=0.49\textwidth]{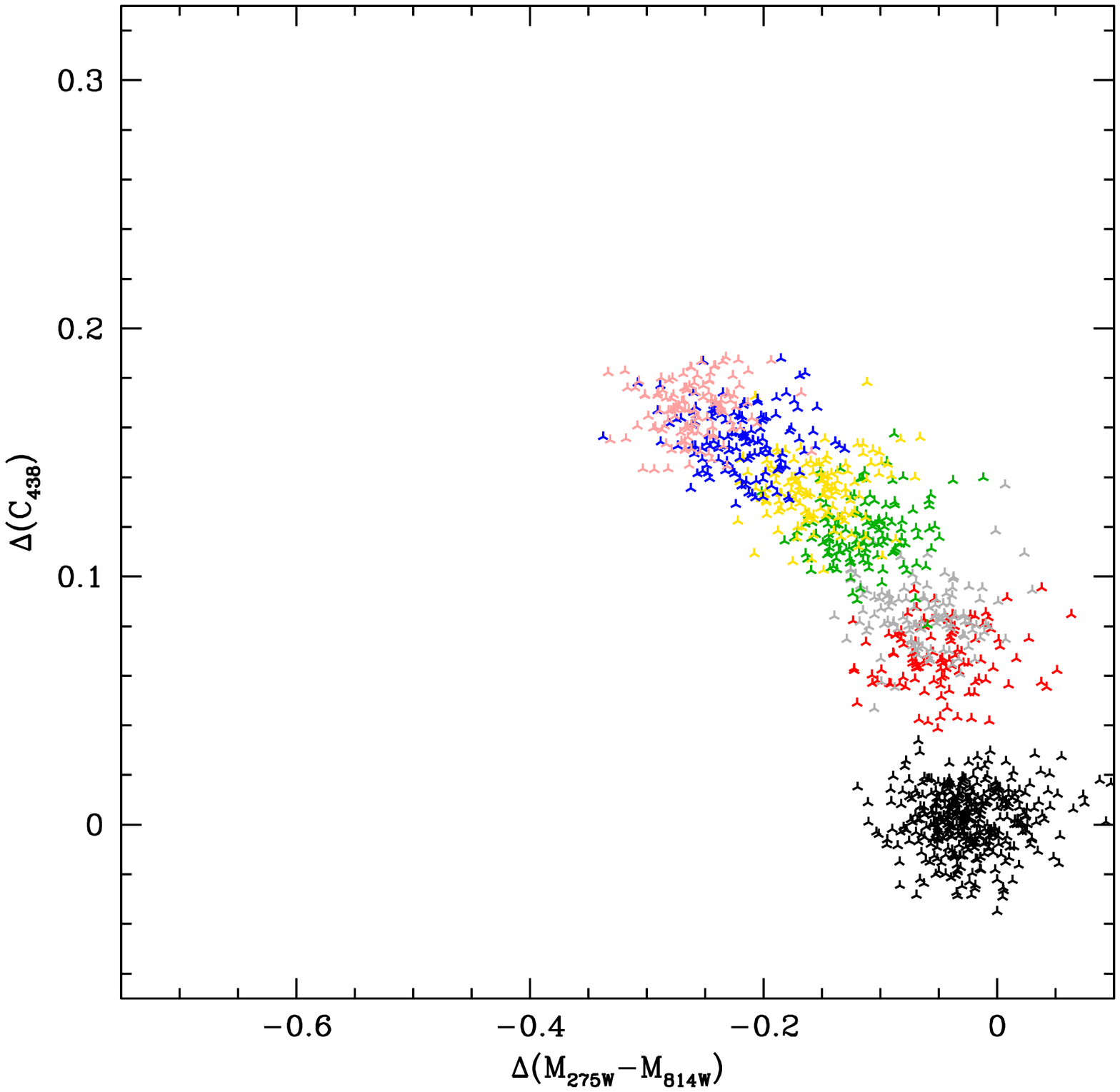}
\caption{Same as Fig.\ \ref{fig_cmd}, including chromospheric emission in the \ion{Mg}{ii} HK lines.}
\label{fig_chromo}
\end{figure*}

\citet{pm11} measured the flux in the \ion{Mg}{ii} HK lines of Galactic giant stars. They concluded that all stars showed chromospheric emission with at least a minimal (basal) flux. However, emission could be higher by about one order of magnitude (see their Fig. 2, as well as \citealt{wood16}). We relied on these studies to estimate the effect of the \ion{Mg}{ii} lines on photometry. For that, we added two emission lines to our synthetic spectra with a Gaussian shape and a total, integrated flux corresponding to the range of values of \citet{pm11}. We subsequently computed photometry and repeated the process for different models along the RGB. We found out that the maximum contribution of \ion{Mg}{ii} lines reaches 0.1 magnitude in the F275W filter. It is achieved for the most luminous stars, because they are also the coolest and thus the ones for which the \ion{Mg}{ii} emission is the strongest, relatively to the photospheric flux. At the bottom of the RGB, the effect of \ion{Mg}{ii} lines is close to zero. This magnitude change implies a color change of the same amount both in  (m$_{275W}$-m$_{814W}$) and C$_{438W}$ since only the F275W filter is affected.

To study the effect of \ion{Mg}{ii} emission on the chromosome map, we first randomly selected a magnitude correction between 0.0 and -0.1. Then, we scaled this correction according to m$_{814W}$, to make sure that the hottest stars have a correction close to zero, and the coolest ones potentially\footnote{Cool stars may also have a negligible correction if their activity level is low.} the maximum correction. Finally, we added this contribution to the photometry of the synthetic cluster shown in Fig.\ \ref{fig_cmd}. The results are shown in Fig.\ \ref{fig_chromo}. The chromosome map and CMDs are almost indistinguishable from those of Fig.\ \ref{fig_cmd}. The reason is that the color corrections caused by \ion{Mg}{ii} lines remain small. Although in the most extreme cases the (m$_{275W}$-m$_{814W}$) and C$_{438W}$ colors could change by 0.1 mag for the coolest/brightest stars, 
this barely happens in our synthetic cluster because; 1) there are few stars at the top of the AGB, and 2) due to random sampling, the corrections even for these stars, remain below 0.1 mag.

We also tested the effect of the \ion{Ca}{ii} lines on the F438W photometry. Assuming the same emission fluxes as for the \ion{Mg}{ii} lines \citep{pm14}, we obtained very small changes ($<$ 0.01 mag). This is because the photospheric flux is much larger in the F438W filter compared to the F275W filter. In addition, the \ion{Ca}{ii} lines are on the very blue side of the filter, where the throughput is small. Thus, these lines do not affect the photometry of GC stars.

\subsection{Continuum emission and variability}

We observe UV emission not only in lines, but also in the continuum \citep{montez17}. The nature of this emission is debated: it may be due to a hot companion \citep{ortiz16} or to heating in the chromosphere, perhaps due to dissipation of acoustic waves or to magnetic reconnections in stars with convective surfaces \citep{schri95}. 
\citet{ortiz16} measured UV flux with \emph{GALEX} in a sample of 58 AGB stars and concluded that 34 of them (i.e., 59\%) showed excess emission below $\sim$2800 \AA\ compared to theoretical SEDs of AGB stars. They argue that this can be a sign of the presence of a companion with \teff\ higher than 5500-6000~K. 

However, the UV flux is usually variable, and some studies report a correlation between UV flux and visual magnitude in AGB stars, favoring a chromospheric origin \citep{sr10,montez17}.
\citet{ortiz19} analyzed the relations between line and continuum UV emission in AGB stars from \emph{GALEX} observations. Continuum emission was evaluated in two bands centered at 3000 and 3200 \AA, while \ion{Mg}{ii} HK lines probed line emission. They showed that the latter varied by a factor $\sim$2 to $\sim$10 with time. The ratio of the fluxes at 3200 and 3000 \AA\ also varied by the same amount and was anticorrelated with line variability: the stronger the \ion{Mg}{ii} HK emission, the smaller the flux ratio F(3200)/F(3000); so, the UV flux is harder when the \ion{Mg}{ii} HK lines are stronger. Finally, they reported that UV emission is dominated by continuum, with lines contributing less than $\sim$36\% of the total flux. Whether these findings also apply to RGB stars is unknown so far. They may be weaker since RGB stars are, on average, hotter than AGB stars, and thus have a stronger photospheric flux in the UV. But one can expect AGB stars in GCs to show UV variability as reported by \citet{ortiz19}. In addition to line emission, one should expect a significant continuum emission resulting from activity.

Whatever the origin of the continuum UV emission of AGB (and potentially RGB) stars, the effect of this emission is similar to that of a companion (see Sect.\ \ref{s_bin}).
A hardening of the UV emission impacts the F275W and F336W filters, and thus CMDs involving these filters, as well as the chromosome map. If binaries produce the UV emission, their fraction among giant stars should be larger than it is among main-sequence stars to reproduce the P1 sequence (Sect.\ \ref{s_bin}). If UV emission is caused by stellar activity, a large fraction of stars should be active.
The main difference is that if chromospheric activity is producing the UV emission, the chromosome map would not be static, in the sense that stars would change position with time according to variability\footnote{Variability may also be present in the case of eclipsing binaries, but these objects should represent only a fraction of the total number of binaries.}. Whether the effect of chromospheric UV emission on the chromosome map is sufficient to quantitatively explain the P1 shape is unclear due to the uncertainties in the knowledge of UV emission of giants.
One may raise an argument against its significance: the clusters M3 and M13 contain stars with presumably very similar properties \citep[metallicity, age;][]{mf09}, but the former shows an extended P1, while the latter does not. If UV variability is an intrinsic property of the stars themselves, and not of the cluster, one does not expect such a different P1 sequence. Clearly, multiepoch observations are needed to investigate the effect of chromospheric activity on MSPs in GCs.

\section{Summary and conclusion}
\label{s_conc}

We presented a study of the effect of binary stars and of chromospheric emission on the shape of the chromosome map of GCs. For that purpose, we first built synthetic clusters using isochrones computed by \citet{wil16}. Along the isochrones, we have computed atmosphere models and synthetic spectra. From the latter, we computed synthetic photometry in the HST filters F275W, F336W, F438W, and F814W. The isochrones in the HRD were thus transposed into CMDs. We subsequently selected different combinations of stars taken from isochrones with various chemical compositions to create a synthetic cluster hosting MSPs. The chromosome map was finally built for this synthetic cluster.

To study the impact of binaries, we first combined the synthetic spectra of stars on the giant branches with those of main-sequence stars. We estimated the changes resulting from the addition of a main-sequence star on the photometry of the giant star. We subsequently used these corrections to replace a fraction of stars from the synthetic cluster by binaries. We then re-built the chromosome map. We proceeded similarly to estimate the effect of chromospheric emission in the \ion{Mg}{ii} HK and \ion{Ca}{ii} HK lines: we added an emission component on top of the stellar spectrum and computed the modified photometry.

We find that binaries contribute to the extension of the P1 sequence. The extension is qualitatively consistent with the observations. However, for the reported binary fractions in GCs of $\lesssim$10\% the number of stars in the extended part of P1 is small. The relative fraction of stars in the extended part and the original sequence is not consistent with observations in NGC~5272, a cluster with properties similar to our synthetic cluster. Even for a larger binary fraction (30\%), the difference remains significant. 
NGC~5272 (M3) and NGC~6205 (M13) are almost twins (same age, mass, metallicity). The former shows an extended P1 while the latter does not, in spite of similar binary fractions.
We thus conclude that while binaries can contribute to the extent of the P1 sequence, they are probably not the main driver, unless binary fractions in GCs are severely underestimated. The minor role of binaries in the extension of P1 is supported by the observations of NGC~6254: it hosts more binaries in its core than in its outer parts, but P1 has the same extension in both regions.

Regarding chromospheric emission, the intensity of \ion{Mg}{ii} HK and \ion{Ca}{ii} HK lines reported in the solar neighborhood are too small to significantly impact the photometry of giant stars. Therefore, the chromospheric emission from these lines does not affect the shape of the chromosome map. Only variations in the UV continuum caused by chromospheric activity could have an effect similar to that of binaries. However, the fraction of stars with significant chromospheric continuum emission is unknown. If multiepoch observations revealed variations in the positions of stars along the P1 sequence, that may be an indication that stellar activity is important in shaping P1 in the chromosome map.
At present, we thus conclude that the extension of the P1 sequence remains enigmatic. 

\section*{Acknowledgments}

We thank an anonymous referee for a prompt and constructive report. This work was supported by the "Programme National de Physique Stellaire" (PNPS) and the "Programme National Cosmologie et Galaxies" (PNCG) of CNRS/INSU co-funded by CEA and CNES. CC acknowledges support from the Swiss National Science Foundation (SNF) for the project 200020-169125 "Globular cluster archeology". WC acknowledges funding from the Swiss National Science Foundation under grant P2GEP2 171971. 
CL thanks the Swiss National Science Foundation for supporting this research under the Ambizione grant number PZ00P2 168065.
FM, CC and WC thank ISSI (International Space Science Institute in Bern, Switzerland) for supporting ISSI Team 271 “Massive Star Clusters across the Hubble Time”.

\bibliographystyle{aa}
\bibliography{chmap_bin}

\end{document}